\documentstyle[12pt]{article}
\title{}
\author{}
\date{}

\begin{document}
\begin{center}
{\large {\bf Energy calculation of magnetohydrodynamic waves and their 
stability for viscous shearing flows.}}
 
\vspace{1cm}
{\bf S.~Chatterjee} {\bf and} {\bf P.~S.~Joarder}

{\it Indian Institute of Astrophysics,

Koramangala, Bangalore -- 560 034, India.}

\vspace{3cm}
ABSTRACT
\end{center}
\noindent A self-consistent, thermodynamic approach is employed to derive 
the {\em wave energy of a magnetohydrodynamic system} within the harmonic 
approximation and to obtain the familiar {\em dispersion relation} from 
the resulting {\em equation of motion}. The evolution of the system
due to an external perturbation is studied by a {\em linear response 
formalism}, that also gives the energy absorbed by the magnetohydrodynamic 
system from the external field. The calculated wave energy reveals the 
presence of {\em positive} and {\em negative energy waves},
that {\em coalesce} together  to give rise to {\em Kelvin - Helmholtz 
instability} of the system.  The threshold value of this instability changes 
only slightly in the presence of a small amount of viscosity, thus precluding 
the {\em dissipative instability} of the negative energy waves. The prediction 
of such a dissipative instability by earlier authors turns out to be the result 
of an erroneous choice of the viscous drag force, that violates the 
fundamental law of {\em Galilean invariance}.

\begin{center}
{\bf 1.~Introduction}
\end{center}

\noindent Propagation of waves in a medium is the manifestation of the 
system's response to a small deviation from its local thermodynamic 
equilibrium. The {\em dispersion relation} $D(k,n)=0$ describes the frequency 
$(n)$ vs. wavenumber $(k)$ relationship of such waves, and the form of the 
{\em dispersion function} $D(k,n)$ is decided by the restoring forces and the 
degrees of freedom that the system possesses.

The degrees of freedom of a magnetohydrodynamic system consist of its density, 
pressure, velocity and its magnetic induction. The conventional method of 
determining the stability of the system and also the nature of its wavemodes, 
is to work with the (MHD) equations connecting the above variables, thus 
obtaining the dispersion relation by requiring a non-trivial solution of the 
problem (eg. Alfven 1950; Cowling 1957; Chandrasekhar 1961).  This method, 
though adequate to study the system's stability, does not directly allow us to 
calculate the energy of the system.

In this paper, we study the response of a magnetohydrodynamic system to 
an external perturbation from the point of view of its energetics. The system
consists of a two-dimensional slab-like inhomogeneity in the plasma and the 
flow parameters of an otherwise uniform magnetic medium, so that the 
equilibrium pressure, density, magnetic field and velocity are given by

\begin{eqnarray}
 p_{\rm 0}(z),~ \rho_{\rm 0}(z), ~ B_{\rm 0}(z), ~ u_{\rm 0}(z) & = &
\left\{ \begin{array}{ll} p_{\rm 0o},~\rho_{\rm 0o},~B_{\rm 0o},~u_{\rm 0o},    & \mbox{$|z|\leq \rm{d},$} \\
p_{\rm 0e},~\rho_{\rm 0e},~B_{\rm 0e},~u_{\rm 0e},    & \mbox{$|z|> \rm{d},$} 
\end{array}
 \right . 
\end{eqnarray}

\noindent with the magnetic field vectors and the steady flows being
aligned to the axis of the slab, ie., to the X-direction.The equilibrium 
condition demands that the plasma inside
and outside the slab are in the total (gas + magnetic) pressure balance,
namely, 
$p_{\rm 0o} + B^{2}_{\rm 0o}/8 \pi  =  p_{\rm 0e} + B^{2}_{\rm 0e}/8 \pi$.
We here note, that Nakariakov and Roberts (1995; see also, Satya Narayanan 
1991; Nakariakov, Roberts and Mann 1996) analysed the normal modes
of the above equilibrium through the solutions of the slab's dispersion 
relation.  These authors also included plasma compressibility in their 
analysis. Considering the algebraic complexity of the calculations of the 
slab's energetics, we here confine ourselves to an {\em incompressible} slab.

In what follows, we begin with the basic definition of the {\em energy density} 
of a magnetohydrodynamic system, deriving  ultimately an expression for the 
space-averged total energy ({\em Hamiltonian}) of the perturbed MHD slab in 
terms of its {\em generalized co-ordinates}, namely, its interfacial 
displacements $\eta(x,z=\pm d,t)$, and their time derivatives $\dot{\eta}
(x,z=\pm d,t)$; see Section 2. In Section 3, we derive the {\em equation of 
motion} of the perturbed slab from the given Hamiltonian, which we solve 
directly to obtain the {\em energy of excitation} of the wavemodes.

The above calculation of the wave energy reveals the presence of {\em negative 
energy waves} (NEW), the existence of which was predicted earlier by various 
authors in both the ordinary hydrodynamic (eg. Benjamin 1963; Cairns 1979; 
Craik and Adam 1979; Ezerskii, Ostrovskii and Stepanyants 1981; Ostrovskii and 
Stepanyants 1982;  Craik 1985; Ostrovski, Rybak and Tsimring  1986) and 
the magnetohydrodynamic (eg. Acheson 1976; Ryutova 1988; Ruderman and Goossens 
1995; Ruderman et al. 1996; Joarder, Nakariakov and Roberts 1997) systems. 
The present paper improves upon these earlier calculations by extending the 
results of Cairns (1979) to the much complicated magnetohydrodynamic 
situations, thus providing a simple expression for the wave energy in terms of 
the linear dispersion function $D(k,n)$ of the MHD slab. In his derivation of 
the wave energy, Cairns followed a procedure that is somewhat similar to the 
ones developed earlier by Stix (1962) and Witham (1974), and used the expansion
$p_{\rm 1,2}\approx D_{\rm 1,2}\left(\omega_{\rm 0}+i\frac{\partial}
{\partial t},k_{\rm 0}\right)A(t)\exp{\left(ik_{\rm 0}x-\omega t\right)}$ 
(see, Cairns 1979) for the pressures across the interface of the fluid.
Such an expansion, though consistent with the dispersion relation, appears to 
us to be intuitively presented by Cairns. Our method, on the other hand, gives 
rigourous calculations of all the physical parameters and, in addition, allows 
one to calculate directly the energy of the system (Section 2), from which 
both the {\em dispersion relation} and the {\em wave energy} follows, 
{\em via.} the equation of motion (see Section 3).
 
Along with many other properties, the negative energy waves are also supposed 
to exihibit {\em dissipative instability}, so that the waves become {\em 
overstable} in the presence of any arbitrarily small amount of viscous 
dissipation in the medium that is in the rest frame of these 
negative energy waves; cf.  Kikina (1967); Weissman (1970); Cairns (1979); 
Ezerskii et al. (1981); Ostrovskii and Stepanyants (1982); Craik (1985); 
Ostrovskii et al. (1986); Ruderman and Goossens (1995); Ruderman et al. (1996).
In this paper, we show that this viscous overstability of the NEW is simply 
the result of an erroneous choice of the dissipative damping, that violates 
the law of {\em Galilean invarience}. In Section 4, we show that both the 
positive and the negative energy waves, in fact, exhibit dissipative damping 
under the action of a viscous drag force, that is consistent with an 
appropriate {\em Galilean transformation}. Concluding remarks are given in 
Section 5, indicating the relevance of this study to several astrophysical MHD 
systems.

\begin{center}
{\bf 2.~Total energy of the perturbed slab}
\end{center}

\noindent 2.1.{\em  The Perturbations:} Consider then, that due to the action 
of some external stress, the interfaces $z=\pm d$ of the magnetic slab 
(see the previous section) are displaced by an amount $\eta (x,z=\pm d,t)=
\tilde{\eta}(k,z=\pm d,t)\exp(ikx)$, where $\tilde{\eta}(k,z=\pm d,t)$ are the 
amplitudes of the Fourier components of the displacements $\eta(x,\pm d,t)$ 
with respect to $x$, and $k$ is the wavenumber. The associated z-displacements 
of the media internal $(|z|\leq d)$ and extrenal $(|z|>d)$ to the slab are 
given by

\begin{eqnarray}
  \eta(x,z,t) & = &  \left\{ \begin{array}{lll}
              \tilde{\eta}(k,z=- d,t)\exp{\left(ikx\right)}\exp{\left\{
k(z+d)\right\}},  & \mbox{$z<- d$,}\\
         \left[\alpha \cosh{\left(kz\right)}+\beta \sinh{\left(kz\right)}
\right] \exp{\left(ikx\right)},  & \mbox{$|z|\leq d$,}\\
              \tilde{\eta}(k,z~=~ d,t)\exp{\left(ikx\right)}\exp{\left\{
- k(z- d)\right\}},   &\mbox{$z>  d$,}
        \end{array}
        \right. 
\end{eqnarray}

\noindent thus showing that for an incompressible plasma slab, as is the case 
considered here, the perturbation amplitudes $\tilde{\eta}(k,z,t)$ are 
evanescent in both the internal ($|z|\leq d$, denoted by `o') and the external 
($|z|>d$, denoted by `e') media away from the interfaces $z=\pm d$ of the 
slab. Such perturbations pertain to the {\it surface modes} of the slab with 
the `cosh' solution in equation (2) presenting a {\it kink (even) surface 
mode}, and the `sinh' solution presenting a {\it sausage (odd) surface mode}; 
cf. Roberts (1981a, b), see also Edwin and Roberts (1982, 1983).

The coefficients $\alpha$ and $\beta$ in equation (2) can be evaluated from 
the condition of continuity of the z-dispacements across the interfaces 
$z~=~\pm d$ of the slab. Retaining the above terminology to reperesent the 
perturbations that are of even or of odd symmetry with respect to the axis
($z=0$) of the magnetic slab, we thus obtain

\begin{eqnarray}
&\left. \begin{array}{ll} \alpha & = \  \tilde{\eta}_{\rm kink}(k,t)/
\cosh{\left(kd\right)},~\tilde{\eta}_{\rm kink}(k,t) = {\frac{1}{2}}
\left\{\tilde{\eta}(k,z=-d,t)+\tilde{\eta}(k,z=d,t)\right\}, \\
\beta & = \  \tilde{\eta}_{\rm sausage}(k,t)/\sinh{\left(kd\right)},
~\tilde{\eta}_{\rm sausage}(k,t) = {\frac{1}{2}}\left\{\tilde{\eta}(k,z=- d,t)
-\tilde{\eta}(k,z=d,t)\right\}. 
\end{array}
\right. &
\end{eqnarray}

So far, we considerd only the z-displacements of the magnetic plasma. The 
perturbations in other magnetohydrodynamic variables can be obtained by 
applying equations (2) and (3) to the linearized MHD 
equations (eg. Chandrasekhar 1961). In medium `o' internal to the slab,
these fluctiations are written as

\newcounter{lett}
\setcounter{lett}{4}
\setcounter{equation}{0}
\renewcommand{\theequation}{\arabic{lett} \alph{equation}}

\begin{eqnarray}
\delta u_{x{\rm o}} & = &-\left(n+ku_{\rm 0o}\right)\left\{\tilde{\eta}_{\rm 
kink}(k,t) \frac{\sinh{\left(kz\right)}}{\cosh{\left(kd\right)}}
+ \tilde{\eta}_{\rm sausage}(k,t)\frac{\cosh{\left(kz\right)}}{\sinh{\left(kd
\right)}}\right\}\exp{\left(ikx\right)}, \\
& & \nonumber \\
\delta u_{z{\rm o}}& = &\left(\frac{n + ku_{\rm 0o}}{n}\right)\left\{\dot
{\tilde{\eta}}_{\rm kink}(k,t)
                 \frac{\cosh{\left(kz\right)}
}{\cosh{\left(kd\right)}}+\dot{\tilde{\eta}}_{\rm sausage}(k,t)
     \frac{\sinh{\left(kz\right)}}{\sinh{\left(kd\right)}}\right\}\exp
{\left(ikx\right)},\\
& & \nonumber \\ 
\delta B_{x{\rm o}} & = &- kB_{\rm 0o}\left\{\tilde{\eta}_{\rm kink}(k,t)
\frac{\sinh{\left(kz\right)}}{\cosh{\left(kd\right)}}
+ \tilde{\eta}_{\rm sausage}(k,t)\frac{\cosh{\left(kz\right)}}{\sinh
{\left(kd\right)}}\right\}\exp{\left(ikx\right)},
\end{eqnarray}

\noindent and

\vspace{-1.0cm}
\begin{eqnarray}
\delta B_{z{\rm o}} & = &\left(\frac{kB_{\rm 0o}}{n}\right)\left\{\dot
{\tilde{\eta}}_{\rm kink}(k,t)
\frac{\cosh{\left(kz\right)}}{\cosh{\left(kd\right)}}+\dot
{\tilde{\eta}}_{\rm sausage}(k,t)
\frac{\sinh{\left(kz\right)}}{\sinh{\left(kd\right)}}\right\}
\exp{\left(ikx\right)}.
\end{eqnarray}

\setcounter{equation}{4}
\renewcommand{\theequation}{\arabic{equation}}

\noindent Here, $\delta u_{x}$ and $\delta u_{z}$ are the perturbations in
the x- and the z- components of the velocity, whereas, $\delta B_{x}$ and 
$\delta B_{z}$ are the perturbations in the x- and the   
z- components of the magnetic field, respectively. In writing equation (4), 
we have used only the $n$-th temporal Fourier mode, so that $\dot{\tilde
{\eta}}(k,t)={\it in}\tilde{\eta}(k,t)$.
 For any arbitrary $\tilde{\eta}(k,t)$, equation (4)
can be generalized by appropriately summing over all possible $n$\ s; such 
an exercise, however, does not alter the results that follow, as can be 
verified by using the condition of independence of the Fourier components.

The fluctuations of the MHD variables in the external ($|z|>d$, denoted by `e'
) media
can similarly be derived. They are

\newcounter{lett1}
\setcounter{lett1}{5}
\setcounter{equation}{0}
\renewcommand{\theequation}{\arabic{lett1} \alph{equation}}

\begin{eqnarray}
\delta u_{x{\rm e}}& = &\mp (n+ku_{\rm 0e})\tilde{\eta}(k,z=\mp d,t)\exp
{\left(ikx\right)}\exp{\left\{\pm k(z\pm d)\right\}}, \\
& & \nonumber \\
\delta u_{z{\rm e}}& = &\left(\frac{n+ku_{\rm 0e}}{n}\right)\dot{\tilde
{\eta}}(k,z=\mp d,t)\exp{\left(ikx\right)}\exp{\left\{\pm k(z\pm d)\right\}}
, \\
& & \nonumber \\
\delta B_{x{\rm e}}& = &\mp\left(kB_{\rm 0e}\right)\tilde{\eta}(k,z=\mp d,t)
\exp{\left(ikx\right)}\exp{\left\{\pm k(z\pm d)\right\}}, 
\end{eqnarray}

\noindent and

\vspace{-1.0cm}
\begin{eqnarray}
\delta B_{z{\rm e}}& = &\left(\frac{kB_{\rm 0e}}{n}\right)
\dot{\tilde{\eta}}(k,z=\mp d,t)
\exp{\left(ikx\right)}\exp{\left\{\pm k(z\pm d)\right\}}. 
\end{eqnarray}

\setcounter{equation}{5}
\renewcommand{\theequation}{\arabic{equation}}

\noindent 2.2. {\em The Energy Densities:} To calculate the total energy 
(or, the {\em Hamiltonian}) of the magnetic slab, we first note that,
the {\em energy density} of an incompressible magnetohydrodynamic system 
is given by
(see the Appendix)

\begin{eqnarray}
\varepsilon & = & p + \frac{B^{2}}{8\pi} + \frac{\rho_{\rm 0}}{2}u^{2},
\end{eqnarray}

\noindent where, $p = p_{\rm 0}+\delta p$, $B = |{\bf B}| = |{\bf B_{\rm 0}} 
+ {\bf {\delta B}}|$,
 and $ u = |{\bf u}| =
|{\bf u_{\rm 0}}+{\bf {\delta u}}|$ are the total pressure, the total 
magnetic 
field strength and the total velocity of the medium, respectively. 
For the specific situation
considered here (see Section 1), we can now use equation (6) to write the 
{\em magnetohydrodynamic
equations of motion} in the following particular form:

\begin{eqnarray}
{\boldmath \nabla}\varepsilon 
& = & - i \rho_{0}\left(n + ku_{0}\right){\boldmath \delta}{\bf u} + 
\frac{ikB_{0}}
{4\pi}{\boldmath \delta}{\bf B}
+\frac{1}{8\pi}{\boldmath \nabla}|{\boldmath \delta}{\bf B}|^2 + 
\rho_{0}u_{0}
{\boldmath \nabla}\left(\delta u_{x}\right),
\end{eqnarray}

\noindent that must be integrated, with the help of equations (4) and (5), 
to obtain the expressions for the energy densities
separately for each of the media internal (`o') and external (`e') to the slab.
 Consider first the internal (`o', $|z|\leq d$) medium.
Equation (4) then allows us to evaluate the various terms on the left hand 
side of equation (7) upto the second order
of smallness in $\tilde{\eta}(k,t)$. Writing $\eta\left(x,t\right)= 
\tilde{\eta}\left(k,t\right)\exp{(ikx)}$, we thus get,

\begin{eqnarray}
{\boldmath \nabla} \varepsilon_{\rm o} & = & 
{\boldmath \nabla}\left[
\frac{{k^2}B_{\rm 0o}^2}{8\pi}\left\{\left[\eta_{\rm kink}(x,t)
\frac{\sinh{\left(kz\right)}}{\cosh{\left(kd\right)}}+\eta_{\rm sausage}(x,t)
\frac{\cosh{\left(kz\right)}}{\sinh{\left(kd\right)}}
\right]^{2} + \right . \right .\nonumber \\
 & & \frac{1}{n^2}\left[\dot{\eta}_{\rm kink}(x,t) \frac{\cosh{\left(kz\right)
}}{\cosh{\left(kd\right)}} \right . + 
\dot{\eta}_{\rm sausage}(x,t)
\left . \left . \frac{\sinh{\left(kz\right)}}{\sinh{\left(kd\right)}} 
\right]^{2} \right\}
 + \nonumber  \\
& & \left\{\rho_{\rm 0o}\frac{\left(n+ku_{\rm 0o}\right)^2}{k}- \right .
\left . \frac{kB_{\rm 0o}^2}{4\pi}\right\}
\left\{\eta_{\rm kink}(x,t) \frac{\sinh{\left(kz\right)}} {\cosh{\left(kd\right
)}} \right . +  \nonumber \\
& & ~~~~~~~~~~~~~~ {\eta}_{\rm sausage}(x,t)
\left . \left . \frac{\cosh{\left(kz\right)}}{\sinh{\left(kd\right)}}\right\}   + \rho_{\rm 0o}u_{\rm 0o}  \left(\delta u_{x{\rm o}}\right)\right],\hspace*{-30ex}
\end{eqnarray}

\noindent which, after integration, gives the following expression for 
the energy density
$\varepsilon_{\rm o}$ of the medium internal to the slab:

\begin{eqnarray}
\varepsilon_{\rm o}  & = &  \left[\rho_{\rm 0o}\frac{\left(n+ku_
{\rm 0o}\right)^2}{k}-
\frac{{kB_{\rm 0o}}^2}{4\pi}\right]\left\{\eta_{\rm kink}(x,t)
\frac{\sinh{\left(kz\right)}}{\cosh{\left(kd\right)}}
+\eta_{\rm sausage}(x,t)\frac{\cosh{\left(kz\right)}}{\sinh{\left(kd\right)}}
\right\} +  \nonumber \\
& &  \frac{{k^2}{B_{\rm 0o}^2}}{8\pi}\left\{\left[\eta_{\rm kink}(x,t)
\frac{\sinh{\left(kz\right)}}{\cosh{\left(kd\right)}}+\eta_{\rm sausage}(x,t)
\frac{\cosh{\left(kz\right)}}{\sinh{\left(kd\right)}}\right]^2 +
\frac{1}{n^2}\left[\dot{\eta}_{\rm kink}(x,t)\frac{\cosh{\left(kz\right)}}
{\cosh{\left(kd\right)}} \right. \right. \nonumber  \\
& & \left. \left. + \dot{\eta}_{\rm sausage}(x,t)\frac{\sinh{\left(kz\right)}}
{\sinh{\left(kd\right)}} \right]^2 \right\} + 
\rho_{\rm 0o} u_{\rm 0o} \left(\delta u_{x{\rm o}}\right) + 
\varepsilon_{\rm 0o} + \delta \varepsilon_{\rm 0o}(x,z,t) + K_{\rm o}. 
\end{eqnarray}

\noindent Here, the quantities $\varepsilon_{\rm 0o}$ and $K_{\rm o}$ are 
constants of integration, that
are independent of $x$ and $z$. The function $\delta \varepsilon_{\rm 0o}(x,z,t
)$ in equation (9) represents some initial 
fluctuations in the equilibrium energy density $\varepsilon_{\rm 0o} = 
p_{\rm 0o} + B_{\rm 0o}^2/8 \pi + \left(\rho_{\rm 0o}/2\right)u_{\rm 0o}^2$ 
of the medium `o',
that are the manifestations of some externally applied initial
stress on the system. Such a stress does not appear explicitly in equation (7), but is 
incorporated phenomenologically in equation (9) to facilitate the study of the 
system's response $\eta(x,z,t)$ to such external perturbations $\delta 
\varepsilon_{\rm 0}$ in Section 3.2 of this paper.

In what follows, we further consider the fluctuations $\delta \varepsilon_{\rm 0}(x,z,t)$ 
to have
only a piece-wise dependence on the $z$-locations; ie., for any particular 
value of $x$, the magnitudes of $\delta \varepsilon_{0}(x,z,t)$ may change 
abruptly across the interfaces $z=\pm d$ of the
slab, but are constant in each of the three media, namely, the medium 
`o' inside $(|z|\leq d)$ the slab, and the medium `e' on either
($z < d$ and $z > d$) side of the slab. For the medium `o', we can write 
$\delta \varepsilon_{\rm 0o}(x,z,t) = \delta \varepsilon_{\rm 0o}(x,z=0,t)$ 
in such a case.

Unlike $\varepsilon_{\rm 0o}$ and $\delta \varepsilon_{\rm 0o}(x,0,t)$, 
the quantity $K_{\rm o}$ in equation (9) contains 
terms, that are of second order smallness in the amplitudes of the 
interfacial displacements $\tilde{\eta}(x,\pm d,t)$. Such second order 
terms do not follow from the {\em linearized} equation (8), 
but are ought to be introduced, as we have to determine all the 
harmonic terms in the system. A general expression
for $K_{\rm o}$ can therefore be written as

\begin{eqnarray}
K_{\rm o} & = & a\ {\tilde{\eta}_{\rm kink}^2}(k,t) + b\ {\tilde{\eta}_
{\rm sausage}^2}(k,t) + c\ {\tilde{\eta}_{\rm kink}}(k,t){\tilde{\eta}_{\rm 
sausage}}(k,t),
\end{eqnarray}

\noindent where, the coefficients $a$, $b$ and $c$ must be determined 
separately from some physical considerations. In order to determine
 these coefficients, it is convenient to first drop the term $\rho_{\rm 0o}
u_{\rm 0o}\left(\delta u_{x{\rm o}}\right)$ from equation (9). Such terms
do not contribute to the total Hamiltonian $H(\tilde{\eta}(k,\pm d,t),
\dot{\tilde{\eta}}(k,\pm d,t))$ of the system, as the integrals like 
$\rho_{\rm 0}u_{\rm 0}\int\!\int \left(\delta u_{x}\right)\,dz\,dx$
represent the velocity of the centre of mass of the respective layers, 
and hence are identically zero owing to wave propagation. Unfortunately, this
property of the integrals has to be imposed and cannot be 
demonstrated here, as the linearization limit that we follow, introduces
some unphysical $\tilde{\eta}^2(k,\pm d,t)$ -type terms in the 
Hamiltonian of the system.  

In order to identify the constants $a$, $b$, $c$ in equation (10), 
we use now the {\em condition of equipartition}. After dropping the term 
involving $\delta u_{x\rm o}$ in equation (9),
we next consider an averaging of the energy density over a time-scale that 
is much longer than any periodicity present in the system. Such
an averaging retains only the second order terms on the right hand side of 
equation (9), while all the terms linear in $\tilde{\eta}(k,\pm d,t)$
drop out. By denoting time averages by overbars, and by using the Parseval 
formula for Fourier transforms, we thus obtain 

\vspace{+1cm}
\begin{eqnarray}
& & 2\overline{\left[\varepsilon_{\rm o}\left(x,z,t\right)- \varepsilon_
{\rm 0o}
-\delta \varepsilon_{\rm 0o}\left(x,0,t\right)\right]}  =  
\left\{\left[\frac {{k^2}B_{\rm 0o}^2}
{8\pi}\frac{\sinh^2{\left(kz\right)}}{\cosh^2{\left(kd\right)}}  + 
a \right]\overline{{\tilde{\eta}_{\rm kink}^2}\left(k,t
\right)} \right.\nonumber  \\
& & ~ + ~ \left .\left[\frac{{k^2}B_{\rm 0o}^2}{8\pi}\frac{\cosh^2{
\left(kz\right)}}
{\sinh^2{\left(kd\right)}}+ b\right]
\overline{{\tilde{\eta}_{\rm sausage}^2}\left(k,t\right)}+ \right.
\left. \frac{{k^2}B_{\rm 0o}^2}{8\pi n^2}\left[\frac{\cosh^2{\left(kz\right)
}}{\cosh^2{\left(kd\right)}}
\overline{\dot{\tilde{\eta}}_{\rm kink}^2\left(k,t\right)}\right. \right . 
 \\
& & \left .\left. + ~ \frac{\sinh^2{\left(kz\right)}}
{\sinh^2{\left(kd\right)}}
\overline{\dot{\tilde{\eta}}
_{\rm sausage}^2\left(k,t\right)}\right]+\right . 
\left. \left[\frac{{k^2}B_{\rm 0o}^2}{4\pi}\frac{\sinh{\left(2kz\right)}}
{\sinh{\left(2kd\right)}}+ c\right]
\overline{\tilde{\eta}_{\rm kink}(k,t)\tilde{\eta}_{\rm sausage}(k,t)} \right. 
\nonumber   \\
& &  \left. + ~\frac{{k^2}B_{\rm 0o}^2}{4\pi}\frac{\sinh{\left(2kz\right)}}
{\sinh{\left(2kd\right)}}\overline{\dot{
\tilde{\eta}}_{\rm kink}(k,t)\dot{\tilde{\eta}}_{\rm sausage}(k,t)}\right\}
.\nonumber
\end{eqnarray}

On the basis of the {\em principle of equipartition of energy} 
in harmonic oscillators, we may now argue, that the contributions
from the terms containing $\tilde{\eta}_{\left(\rm kink/sausage\right)}^2
(k,t)$ must be equal to the contributions from the terms
containing $\dot{\tilde{\eta}}_{\left(\rm kink/sausage\right)}^2(k,t)$. 
Applying similar arguments for the cross terms (containg $\tilde{\eta}_
{\rm kink}(k,t)\tilde{\eta}_{\rm sausage}(k,t)$ etc.), we ultimately get

\begin{eqnarray}
a & = & \frac{{k^2}B_{\rm 0o}^2}{8\pi}\frac{1}{\cosh^2{\left(kd\right)}},
\ b = - \frac{{k^2}B_{\rm 0o}^2}{8\pi}\frac{1}{\sinh^2{\left(kd\right)}},\  
c = 0.
\end{eqnarray}

Using equations (9-12), we finally arrive at the expression for the energy
density of the medium `o' internal to the slab. This expression reads

\begin{eqnarray}
& & \varepsilon_{\rm o}(x,z,t)  =  \varepsilon_{\rm 0o} + 
\delta \varepsilon_{\rm 0o}(x,0,t)
+ \frac{{k^2}B_{\rm 0o}^2}{8\pi}\left\{\left[\eta_{\rm kink}^2(x,t) 
\frac{\sinh^2{\left(kz\right)}}{\cosh^2{\left(kd\right)}} \right . \right 
. \nonumber \\
& & \left .~~ + ~~ \frac{1}{2} \frac{\eta_{\rm kink}^2(x,t)}{\cosh^2{(kd)}}
\right]
 \left. + \left[\eta_{\rm sausage}^2(x,t)\frac{\cosh^2{\left(kz\right)}}
{\sinh^2{\left(kd\right)}}-\frac{1}{2}
\frac{\eta_{\rm sausage}^2(x,t)}{\sinh^2{(kd)}}\right] \right.  \\
& & \left. ~~ + ~~ 2\eta_{\rm kink}(x,t)\eta_{\rm sausage}(x,t)
\frac{\sinh{\left(2kz\right)}}{\sinh{\left(2kd\right)}}\right\}
+ \frac{{k^2}B_{0o}^2}{8\pi\rm n^2}\left[\dot{\eta}_{\rm kink}^2(x,t)
\frac{\cosh^2{\left(
kz\right)}}{\cosh^2{\left(kd\right)}}\right. \nonumber \\
& & \left. ~~ + ~~ \dot{\eta}_{\rm sausage}^2(x,t)
\frac{\sinh^2{\left(kz\right)}}{\sinh^2{\left(kd\right)}}+ 
2\dot{\eta}_{\rm kink}(x,t)\dot{\eta}_{\rm sausage}(x,t)\frac{\sinh
{\left(2kz\right)}}
{\sinh{\left(2kd\right)}}\right]\nonumber \\
& & ~~ +  ~~ \left[\rho_{\rm 0o}\frac{\left (n+ku_{\rm 0o}\right)^2}{k} 
- \frac{kB_{\rm 0o}^2}{4\pi}\right]\left\{\eta_{\rm kink}(x,t)
\frac{\sinh{\left(kz\right)}}{\cosh{\left(kd\right)}}
+ \eta_{\rm sausage}(x,t)\frac{\cosh{\left(kz\right)}}{\sinh{\left(kd\right)}}
\right\}
. \nonumber 
\end{eqnarray}

The energy density of the external media can be derived by following a 
similar procedure. This energy density is given
by

\vspace{-0.5cm}
\begin{eqnarray}
& & \varepsilon_{\rm e}(x,z,t)  =  
\frac{k^2B_{0e}^2}{8\pi}\left[\eta^2(x,\mp d,t) + \frac{1}{n^2}\ddot{\eta}
^2(x,\mp d,t)\right]\exp{\left\{\pm 2
k\left(z\pm d\right)\right\}}\nonumber  \\
& & ~~~  \mp ~~~ \left[\frac{kB_{\rm 0e}^2}{4\pi}-\rho_{\rm 0e}
\frac{\left(n + ku_{\rm 0e}\right)^2}{k}\right]\eta(x,\mp d,t)
{\left\{\pm k\left(z \pm d\right)\right\}} \nonumber \\
& & ~~~ +  ~~~ \varepsilon_{\rm 0e}+\delta \varepsilon_{\rm 0e}(x,\mp d,t) ,
\end{eqnarray}

\noindent where, the `-' and the `+' signs denote the regions 
$z <-d$ and $z>d$, respectively.
In writing equation (14), we have assumed that some initial fluctuations, 
in the form of $\delta \varepsilon_{\rm 0e}(x,\mp d,t)$
, is imposed on the equilibrium state $\varepsilon_{\rm 0e}$ of 
the external plasma on both the sides of the magnetic
slab, that are the manifestations of some external stress applied on 
the system; see the discussions following equation (9) above.

\noindent 2.3. {\em The Total Hamiltonian:} The total energy, or the 
Hamiltonian 
$H(\tilde{\eta},\dot{\tilde{\eta}})$ of the given MHD system
is derived by integrating its energy density
over the entire
volume of the system consisting of the magnetic slab in its magnetic 
environment. Thus

\begin{eqnarray}
H\left(\tilde{\eta}(k,z=\pm d,t),\dot{\tilde{\eta}}(k,z=\pm d,t)\right)  & =&  
\int\!\! \int_{-\infty}^{\infty} \kern -.2in 
\varepsilon(x,z,t) \,dz\,dx 
\nonumber \\
  = \int\!\left\{\int
_{-\infty}^{-d+\tilde{\eta}(k,z=-d,t)\exp{\left(ikx\right)}}\kern -1.4in 
\varepsilon_{e}(x,z,t)\,dz \kern 0.5in  \right .
&   + &   \int_{-d+\tilde{\eta}(k,z=-d,t)\exp{\left(ikx\right)}} 
^{d+\tilde{\eta}(k,z=d,t)\exp{\left(ikx\right)}}   \kern -1.3in
 \varepsilon_{o}(x,z,t)\,dz 
\kern .4in  \nonumber \\
& & + \left .\int_{d+\tilde{\eta}(k,z= d,t) 
\exp{\left(ikx\right)}}^{\infty}\kern -1.1in \varepsilon_{e} 
(x,z,t)\,dz\kern .3in \right\}\!dx,
\end{eqnarray}

\noindent where, the quantities $\varepsilon_{\rm o}(x,z,t)$ and 
$\varepsilon_{\rm e}(x,z,t)$ are as given in equations (13) and (14), 
respectively.

After performing the integrations as indicated in equation (15), 
while retaining terms, that are of second order of smallness in the 
displacements $\eta(x,z,t)$, we finally arrive at the required expression 
for the total energy of the perturbed magnetic slab.  This expression reads

\vspace{-0.5cm}
\begin{eqnarray}
& & H\left(\tilde{\eta}_{\rm kink},\tilde{\eta}_{\rm sausage},
\dot{\tilde{\eta}}_{\rm kink},\dot{\tilde{\eta}}_{\rm sausage}
\right) = - \tilde{\sigma}_{\rm kink}(-k,t)\tilde{\eta}_{\rm kink}(k,t) 
- \tilde{\sigma}_{\rm sausage}(-k,t)\tilde{\eta}_{\rm sausage}
(k,t) \nonumber \\
& & ~~~ + ~~~ \tilde{\eta}_{\rm kink}^2(k,t)\left\{\left[\rho_{\rm 0e}
\frac{\left(n+ku_{\rm 0e}\right)^2}{k}-
\frac{3kB_{\rm 0e}^2}{16\pi}\right] \right. 
\left. + \left[\rho_{\rm 0o}\frac{\left(n+ku_{\rm 0o}\right)^2}{k}-
\frac{kB_{\rm 0o}^2}{4\pi}\right]\tanh{\left(
kd\right)}\right. \nonumber  \\
& & \left. ~~~ + ~~~ \frac{{k^2}B_{\rm 0o}^2}{8\pi}\left[\frac{2\sinh
{\left(2kd\right)}+4\left(kd\right)}{8k\cosh^2\left(kd\right)}
\right]\right\} + 
 \tilde{\eta}_{\rm sausage}^2(k,t)\left\{\left[\rho_{\rm 0e}
\frac{\left(n+ku_{\rm 0e}\right)^2}{k}
- \frac{3kB_{\rm 0e}^2}{16\pi}\right] \right. \nonumber \\
& & \left. ~~~ + ~~~ \left[\rho_{\rm 0o}\frac{\left(n+ku_{\rm 0o}\right)^2}{k}
-\frac{kB_{\rm 0o}^2}{4\pi}\right]
\coth{\left(kd\right)} \right. 
\left. + \frac{kB_{\rm 0o}^2}{8\pi}\left[\frac{2\sinh{\left(2kd\right)}-
4\left(kd\right)}
{8k\sinh^2\left(kd\right)}\right]\right\}  \nonumber \\
& & ~~~ + ~~~ \dot{\tilde{\eta}}_{\rm kink}^2(k,t)\left\{\frac{kB_{\rm 0e}^2}
{16\pi{n^2}} + \frac{k^2B_{\rm 0o}^2}{8\pi n^2}
\left[\frac{2\sinh{\left(2kd\right)}+4\left(kd\right)}{8k\cosh^2{\left
(kd\right)}}\right]\right\} \\
& & ~~~ + ~~~ \dot{\tilde{\eta}}_{\rm sausage}^2(k,t)\left\{\frac{kB_
{\rm 0e}^2}{16\pi{n^2}} + \frac{k^2B_{\rm 0o}^2}{8\pi n^2}
\left[\frac{2\sinh{\left(2kd\right)}-4\left(kd\right)}{8k\sinh^2{\left
(kd\right)}}\right]\right\}\nonumber, 
\end{eqnarray}

\newpage
\noindent where, we have used the definitions

\newcounter{lett2}
\setcounter{lett2}{17}
\setcounter{equation}{0}
\renewcommand{\theequation}{\arabic{lett2} \alph{equation}}

\begin{eqnarray}
\delta \varepsilon_{\rm 0e}(x,-d,t)-\delta \varepsilon_{\rm 0e}(x,d,t) & 
= & \sum_{k^{\prime}} 
\tilde{\sigma}_{\rm kink}\left(k^{\prime},t\right)\exp{\left(ik^{\prime}x
\right)},
\end{eqnarray}

\noindent and

\vspace{-1.0cm}
\begin{eqnarray}
\delta \varepsilon_{\rm 0e}(x,-d,t)+\delta \varepsilon_{\rm 0e}(x,d,t)-
2\delta \varepsilon_{\rm 0o}(x,0,t)
& = & \sum_{k^{\prime}}\tilde{\sigma}_{\rm sausage}\left(k^{\prime},t\right)
\exp{\left(
ik^{\prime}x\right)},
\end{eqnarray}

\setcounter{equation}{17}
\renewcommand{\theequation}{\arabic{equation}}

\noindent with $\tilde{\sigma}_{\rm kink}\left(k^{\prime},t\right)$ and 
$\tilde{\sigma}_{\rm sausage}\left(k^{\prime},t\right)$ being the $k^\prime$
-th Fourier amplitudes of the externally applied stresses, that excite the 
{\em kink (even)} and the {\em sausage (odd)} perturbations of the slab, 
respectively.

\begin{center}
{\bf 3.~Excitation energy of the wavemodes.}
\end{center}

\noindent 3.1. {\em Equation of Motion and the Dispersion Relations:} The 
Hamiltonian, that we  derive in the previous section (see equation (16)), 
immeditaely allows us to write down the {\em equation of motion} for the 
perturbed magnetic slab. This equation reads

\begin{eqnarray}
& &\left\{\frac{kB_{\rm 0e}^2}{8\pi n^2}+\frac{k^2B_{\rm 0o}^2}{8\pi n^2}
\left[
\frac{2\sinh{\left(2kd\right)}\pm4\left(kd\right)}{4k\cosh^2{\left(kd\right)}}
\right]\right\}\ddot{\tilde{\eta}}_{\left(\rm kink/sausage\right)}(k,t) 
\nonumber \\
& & +2\left\{\left[\rho_{\rm 0e}\frac{\left(n+ku_{\rm 0e}\right)^2}{k}-
\frac{3kB_{\rm 0e}^2}{16\pi}\right] + 
\left[\rho_{\rm 0o}\frac{\left(n + ku_{\rm 0o}\right)^2}{k}-\frac{kB_{\rm 0o}
^2}{4\pi}\right]f(k)
\right. \nonumber \\
& & + \left. \frac{k^2B_{\rm 0o}^2}{8\pi}\left[\frac{2\sinh\left(2kd\right)
\pm4kd}{8k\cosh^2{\left(kd\right)}}\right]\right\}\tilde{
\eta}_{\left(\rm kink/sausage\right)}(k,t)\nonumber \\
& &= \tilde{\sigma}_{\left(\rm kink/sausage\right)}(-k,t),
\end{eqnarray}

\noindent where, the `+' and the `-' sign applies to the cases of the 
{\em kink} (with $f(k)=\tanh{(kd)}$) and the {\em sausage }(with $f(k)=
\coth{(kd)}$) perturbations, respectively.

To check the correctness of equations (16) and (18), we consider, for the 
time being, that there is no external stress applied on the system, ie., 
$\tilde{\sigma}(-k,t)= 0$. The motion of the plasma slab then consists of 
its various eigenmodes, as revealed by its {\em dispersion relations} 
$D(k,n)= 0$, connecting the frequency ($n$) and the wavenumber ($k$) of any 
temporal Fourier component $\overline{\eta}(k,n)\exp{(int)}$ of the 
displacements $\tilde{\eta}(k,t)$ of the slab.
For the kink and the sausage disturbances, these dispersion relations are 
obtained by substituting $\partial/\partial t = in$ in equation (18). For 
arbitrary $\overline{\eta}(k,t)$, we thus obtain

\begin{eqnarray}
D_{\left(\rm kink/sausage\right)}(k,n)  =  \rho_{\rm 0e}\left[\left(n+ku_
{\rm 0e}\right)^2-k^2c_{\rm Ae}^2\right] + \rho_{\rm 0o}\left[\left(n+ku_
{\rm 0o}
\right)^2-k^2c_{\rm Ao}^2\right]f\left(k\right) = 0,
\end{eqnarray}

\noindent where, we have introduced the {\em Alfven} speeds $c_{\rm Ao} = 
B_{\rm 0o}/{\left(4\pi\rho_{\rm 0o}\right)}^{1/2}$ and $c_{\rm Ae} = 
B_{\rm 0e}/{\left(4\pi\rho_{\rm 0e}\right)}^{1/2}$ in the two media `o' and 
`e', respectively.

We here note, that equation (19)
is essentially the same as the dispersion relations obtained by Nakariakov and 
Roberts (1995), for the normal modes of an incompressible magnetic slab 
embedded in an incompresible MHD medium, where there is a relative tangential 
velocity between the slab and its environment. This equation admits solutions

\newcounter{lett3}
\setcounter{lett3}{20}
\setcounter{equation}{0}
\renewcommand{\theequation}{\arabic{lett3} \alph{equation}}

\begin{eqnarray}
n_{\pm} & = & - k\left[\left\{\alpha_{{\rm e}f}u_{\rm 0e}+\alpha_{{\rm o}f}
u_{\rm 0o}\right\}
\pm \left\{c_{{\rm k}f}^2 - \alpha_{{\rm o}f}\alpha_{{\rm e}f}
\left(u_{\rm oe}- u_{\rm 0o}\right)^2\right\}^{1/2}
\right],
\end{eqnarray}

\noindent with

\vspace{-1.0cm}
\begin{eqnarray}
\alpha_{{\rm e}f}=\frac{\rho_{\rm 0e}}{\rho_{\rm 0e}+\rho_{\rm 0o}f(k)}, 
\alpha_{{\rm o}f}
& = &  \frac{\rho_{\rm 0o}f(k)}{\rho_{\rm 0e}+\rho_{\rm 0o}f(k)},\mbox {and}
\  c_{{\rm k}f}^2 =
\frac{\rho_{\rm 0e}c_{\rm Ae}^2+\rho_{\rm 0o}f(k)c_{\rm Ao}^2}{\rho_{\rm 0e}+
\rho_{\rm 0o}f(k)}.
\end{eqnarray}

\setcounter{equation}{20}
\renewcommand{\theequation}{\arabic{equation}}

Each of the {\em kink} ($f(k)=\tanh{\left(kd\right)}$) and the {\em sausage} 
($f(k)=\coth{\left(kd\right)}$) solutions then allows two distinct eigenmodes 
of the slab, that are represented by the `+' and the `-' signs in equation 
(20). These modes are purely oscillatory (giving {\em surface waves}) when the 
discriminant of equation (20) is real, but one of them (the `+' mode) becomes 
a growing mode (giving {\em Kelvin-Helmholtz (K-H) instability}) when this 
discriminant is imaginary. The instability thus sets in for relative velocities

\begin{eqnarray}
|u_{\rm 0e}-u_{\rm 0o}| & \geq & c_{{\rm k}f}/\left(\alpha_{{\rm o}f}
\alpha_{{\rm e}f}\right)^{1/2},
\end{eqnarray}

\noindent ie., for values at which the `+' and the `-' surface modes of either 
the kink or the sausage- type {\em coalesce} together in the real $n$ vs. $k$ 
plane to produce an unsatble region of complex frequency; see Cairns (1979) 
for some examples of such {\em coalescence instabilities} drawn from 
hydrodynamics. Such a coalescence instability occurs only when the 
modes involved have energies of opposite sign; see the discussions in the 
next section. We also note, that equation (21) reduces to the instability 
criterion given by Singh and Talwar (1994) and Nakariakov and Roberts (1995) 
in a situation, where the plasma slab moves in a static environment, ie., when
$u_{0e}= 0$.

Equation (21) further shows, that the kink and
the sausage modes merge together to give only two (`+' and `-') surface modes 
in the case of an infinitely wide ($kd \rightarrow \infty,\ f(k) \rightarrow 
1$) slab, so that the quantity $c_{{\rm k}f}$ coincides
with the {\em phase speed} $c_{\rm k}= {\left(\rho_{\rm 0o}c_{\rm Ao}^2 + 
\rho_{\rm 0e}c_{\rm Ae}^2\right)}^{1/2}/\left
(\rho_{\rm 0o}+\rho_{\rm 0e}\right)^{1/2}$ {\em of the hydromagnetic 
surface waves} (Roberts 1981a,b)
in a single surface of discontinuity separating two uniform magnetic
plasma media. The instability criterion
reduces to the classical threshold for the K-H instability of a magnetic
tangential discontinuity (Chandrasekhar 1961) in this limit of an infinitely 
thick plasma slab.

\noindent 3.2. {\em Evolution of the System under External Stress:} With this 
brief discussion on the nature of the normal modes of the slab, we now turn
to the calculations for the work done by an external stress $\tilde{\sigma}
(-k,t)$ in exciting each of these modes, which, in turn, is stored as the 
energy of that particular wavemode of the system. To find
this wave energy, we consider that the external stresses begin to act on the 
system at a time $t=0$, so that

\begin{eqnarray}
\tilde{\sigma}_{\left(\rm kink/sausage\right)}\left(-k,t\right) & = & 
\tilde{\sigma}_{\left(\rm kink/sausage\right)}(-k)
\Theta(t)F(t),
\end{eqnarray}

\noindent with $F(t)$ being an arbitrary function of time $t$, and 
$\Theta(t)$ being a {\em Heaviside unit step function}. The {\em causal 
response} of the system to this external stress is given in terms
of a {\em response function} $G(k,t)$. Thus

\begin{eqnarray}
\tilde{\eta}_{\left(\rm kink/sausage\right)}(k,t) & = & \left\{ 
\begin{array}{ll}
                  0   & \mbox{ if $t<0$,} \\
{\displaystyle \int}_{0}^{t} {G_{\left(\rm kink/sausage\right)}
\left(k,t- t^{\prime}
\right)F}(t^{\prime})\,dt^{\prime} & \mbox{ if $t\geq 0$,}
\end{array}
\right.
\end{eqnarray}

\noindent the {\em Laplace transform} of which is given by

\begin{eqnarray}
\tilde{\eta}^L_{\left(\rm kink/sausage\right)}(k,s) & = & G^L_{
\left(\rm kink/sausage\right)}(k,s)
F^{L}(s),
\end{eqnarray}

\noindent with the superscript `L' denoting a Laplace transform.
The one-sided Fourier transform $\overline{\eta}\left(k,\omega\right)$ of 
$\tilde{\eta}\left(k,t>0\right)$ can now
be found from equation (24), by analytically continuing $s$ to $i\omega$, 
so that

\begin{eqnarray}
\overline{\eta}_{\left(\rm kink/sausage\right)}(k,\omega) & = & 
\tilde{\eta}^L_{\left(\rm kink/sausage\right)}(k,i\omega) = G^L_{
\left(\rm kink/sausage\right)}(k,i\omega)F^L(i\omega).
\end{eqnarray}

\noindent With the help of equations (19) and (25), equation (18) yields

\begin{eqnarray}
D_{\left(\rm kink/sausage\right)}(k,\omega)G^L_{\left(\rm kink/sausage\right)}
(k,i\omega)F^L(i\omega)& = & k\tilde{\sigma}_{\left(\rm kink/sausage\right)}
(-k)F^L(i\omega),
\end{eqnarray}

\noindent whence we obtain
\begin{eqnarray}
G^L_{\left(\rm kink/sausage\right)}(k,i\omega) = k\tilde{\sigma}_{
\left(\rm kink/sausage\right)}
(-k)/D_{\left(\rm kink/sausage\right)}(k,\omega).
\end{eqnarray}

\noindent Equation (27) helps us to find $G^L(k,s)$ for any complex value of s, 
by means of analytic continuation. Thus

\begin{eqnarray}
G^L_{\left(\rm kink/sausage\right)}(k,s) & = & k\tilde{\sigma}_{\left(\rm 
kink/sausage\right)}(-k)/D_{\left(\rm kink/
sausage\right)}(k,- is),
\end{eqnarray}

\noindent which, combined with equation (24), ultimately gives the expression 
for the Laplace transform $\tilde{\eta}^L_{\left(\rm kink/sausage\right)}
(k,s)$ of the displacements $\tilde{\eta}_{\left(\rm kink/sausage\right)}
(k,t)$ for any complex value of s. This expression is given
as

\begin{eqnarray}
\tilde{\eta}^L_{\left(\rm kink/sausage\right)}(k,s) & = & k 
\tilde{\sigma}_{\left(\rm kink/sausage\right)}(-k)F^L(s)/D_{\left(\rm 
kink/sausage\right)}(k,-is),
\end{eqnarray}

\noindent from which we can find the time evolution of the displacements 
$\tilde{\eta}_{\left(\rm kink/sausage\right)}(k,t)$ by means of {\em 
Bromwich's integral formula}. Thus,

\begin{eqnarray}
\tilde{\eta}_{\left(\rm kink/sausage\right)}(k,t)  =  \frac{k
\tilde{\sigma}_{\left(\rm kink/sausage\right)}(-k)}{2\pi i}
{\displaystyle \int}_{\Upsilon - i\infty}^{\Upsilon + i\infty}F^L(s)
\exp{\left(st\right)}\,ds /D_{\left(\rm kink/sausage\right)}(k,-is),
\end{eqnarray}

\noindent where, the real constant $\Upsilon$ is so chosen that the 
singularities of the integrand lie to the left of the line $s=\Upsilon$, the 
singularities themselves being of the nature of simple poles, that are the 
solutions of the dispersion relations $D_{\left(\rm kink/sausage\right)}(k,n)= 
0$. In the present case, these singularities are $s_{\pm}=in_{\pm}$, with 
$n_{\pm}$ being simply the frequencies given in equation (20) above.

To evaluate the integral given in equation (30), we need now to consider
some particular form of the function $F(t)$ for the time dependent part of the 
external stress $\tilde{\sigma}_{\left(\rm kink/sausage\right)}(-k,t)$; see,
equation (22). A common model to choose is an exponential one, so that

\begin{eqnarray}
F(t) & = & A\exp{\left(-\lambda t\right)},
\end{eqnarray}

\noindent with A being a constant, and the exponent $\lambda $ being positive 
definite. For this particular functional form of the external stress 
$\tilde{\sigma}(-k,t)$, equation (30) yields

\begin{eqnarray}
& & \tilde{\eta}_{\left(\rm kink/sausage\right)}(k,t) = \frac{k
\tilde{\sigma}_{\left(\rm kink/sausage\right)}(-k)}{2\pi i}
A{\displaystyle \int}_{\Upsilon - \infty}^{\Upsilon + \infty}\frac{\exp{
\left(st\right)}\,ds}{D_{\left(\rm kink/sausage\right)}
(k,-is)}\frac{1}{s+\lambda} \nonumber \\
& & ~~~ = ~~~~~ k\tilde{\sigma}_{\left(\rm kink/sausage\right)}(-k)A
\left[\frac{\exp{\left(-\lambda t\right)}}{D_{\left(\rm kink/sausage\right)}
(k,-\lambda)} + \frac{\exp{\left(s_{+}t\right)}}{\left[\frac{\partial D_{
\left(\rm kink/sausage\right)}(k,-is)}{\partial s}\right]_{s_{+}}}
\frac{1}{s_{+}+\lambda} \right . \nonumber \\
& & \left . ~~~~~~~~~~ + \frac{\exp{\left(s_{-}t\right)}}{\left[\frac{\partial 
D_{\left(\rm kink/sausage\right)}(k,-is)}{\partial s}\right]_{s_{-}}}
\frac{1}{s_{-}+\lambda}\right], 
\end{eqnarray}

\noindent where, the second expression follows from the {\em residue theorem}. 
The first term in this expression is a transient term, that decays with the 
decay of the extrenal stress, whereas, the last two terms are due to 
disturbances that live on even after the withdrawal of the external force and, 
therefore, pertain to the eigenmode oscillations of the slab. For the sake of 
illustrations, we may consider a {\em delta function} type external stressing, 
so that $\lambda \rightarrow \infty$, and $A/\lambda \rightarrow 1$.  The 
displacements $\tilde{\eta}(k,t)$ then evolve as

\begin{eqnarray}
\tilde{\eta}_{\left(\rm kink/sausage\right)}(k,t) & = & {\it i}k
\tilde{\sigma}_{\left(\rm kink/sausage\right)}(-k) \left[\frac{\exp{(in_{+}t)}}
{\left[\frac{\partial D_{\left(\rm kink/sausage\right)}(k,n)}{\partial n}
\right]_{n _{+}}}\right.  \nonumber \\
& & ~~~~~~~~~~~~~~~~~~~~~~~~~~~  \left. + \frac{\exp{(in_{-}t)}}{
\left[\frac{\partial D_{\left(\rm kink/sausage\right)}(k,n)}{\partial n}
\right] _{n_{-}}}\right] \\
& = &\frac{ik\tilde{\sigma}_{\left(\rm kink/sausage\right)}(-k)}{
\left(n_{+}-n_{-}\right)}\left[\exp{(in_{+}t)} -\exp{(in_{-}t)}\right], 
                                                 \nonumber
\end{eqnarray}

\noindent givng us the familiar result, that the application of an 
instantaneous external stress $\tilde{\sigma}(-k)\delta(t)$ on the MHD slab 
creates long lived excitations pertaining to the normal surface modes of 
the slab, that have frequencies $n_{+}$ and $n_{-}$ as given by the dispersion 
relations (20). Equation (33) further shows that the two surface modes,
denoted here by a `+' and a `-' sign, are out of phase with the external 
stress $\tilde{\sigma} (-k,t)$ by $\pi/2$ and $3\pi/2$, respectively. The 
modes are of same amplitude, ie., $|\tilde{\eta}^+_{\left(\rm 
kink/sausage\right)}(k,t)|=|\tilde{\eta}^-_{\left(\rm kink/sausage\right)}
(k,t)|$, but their different phase relations with respect to $\tilde{\sigma}
(-k,t)$ lead to a difference in their respective energy absorption rates, 
as will be presented in the next section.

\noindent 3.3. {\em Energy Absorbed by the Modes:} Equation (33) shows that 
the rate of absorption of energy per unit area by each of the $\tilde{\eta}
^{\pm}_{\left(\rm kink/sausage\right)}(k,t)$ mode, from the external stress 
$\tilde{\sigma}_{\left(\rm kink/sausage\right)}(-k,t)$ is given by

\begin{eqnarray}
& & \dot{h}^{\pm}\left(\tilde{\eta}_{\left(\rm kink/sausage\right)}(k,t),
\dot{\tilde{\eta}}_{\left(\rm kink/sausage\right)}(k,t)\right) = 
\tilde{\sigma}_{\left(\rm kink/sausage\right)}(-k,t)
\dot{\tilde{\eta}}^{\pm}_{\left(\rm kink/sausage\right)}(k,t) \nonumber \\
& & ~~~~~~~~~~~ = ~~~~ - \frac{kA^2|\tilde{\sigma}_{\left(\rm kink/sausage
\right)}(-k)|^2}{\left[\frac{\partial D_{\left(\rm kink/sausage\right)}(k,n)}
{\partial n}\right]_{n_{\pm}}}\frac{n_{\pm}}{in_{\pm}+\lambda}\exp{
\left\{(in_{\pm}-\lambda)t\right\}},
\end{eqnarray}

\noindent whence we calculate the total energy (per unit area) absorbed by the
mode, by integrating equation (34) over a time t that is much longer than the
decay time of the stress, ie. $t \gg 1/\lambda$. This absorption is given by

\begin{eqnarray}
\Delta h_{\left(\rm kink/sausage\right)}(n_{\pm}) & = & {\displaystyle 
\int}_{0}^{\infty} \dot{h}^{(\pm)}\left(\tilde{\eta}_{\left(\rm kink/sausage
\right)}(k,t), \dot{\tilde{\eta}}_{\left(\rm kink/sausage\right)}(k,t)\right)
\,dt \nonumber \\ & = & \frac{k|\tilde{\sigma}_{\left(\rm kink/sausage\right)}
(-k)|^2n_{\pm}}{\left[\frac{\partial D_{\left(\rm kink/sausage\right)}}{
\partial n}\right]_{n_{\pm}}}\frac{A^2}{n_{\pm}^2 + \lambda^2},
\end{eqnarray}

\noindent which, after substitution in favour of the modal amplitude 
$\tilde{\eta}^{(\pm)}_{\left(\rm kink/sausage\right)}(k,t)$ in equation 
(32), ultimately yields

\begin{eqnarray}
\Delta h_{\left(\rm kink/sausage\right)}(n_{\pm}) 
& = & \left(n\frac{\partial D_{\left(\rm kink/sausage\right)}(k,n)}
{\partial n}\right)_{n_{\pm}}\frac{1}{k}|\eta^{(\pm)}_{\left(\rm 
kink/sausage\right)}(k,t)|^2.
\end{eqnarray}

We here note that, inspite of the complexities presented by the magnetic field, the expression of wave energy presented in equation (36) 
is essentially the same as the one given in Cairns (1979) in the case of a 
purely hydrodynamic system.  Unlike Cairns (1979), whose method was intuitive 
(see Section 1 for details), we here derive our results {\em directly} from 
the equation of motion (18) of the system. The particular models of external 
stress that we assume for the purpose of demonstrations, do not have any 
bearing on our final result in equation (36), thus implying that this 
expression for the wave energy is truely a generalized expression for any 
hydrodynamic or magnetohydrodynamic system.

\noindent 3.4. {\em Waves of Negative Energy} (NEW): Consider, for 
simplicity's sake, a frame of reference co-moving with the external medium, so 
that $u_{\rm 0e}=0$. Consider further, that the velocity of the slab 
$u_{\rm 0o}$ is increased gradually from zero through positive values in this 
frame of reference. Equation (20) in Section 3.1 then shows that, as long as 
$u_{0{\rm o}} < c_{{\rm k}f}/\left(\alpha_{{\rm o}f}\right)^{1/2}$, both the 
`+' and the `-' modes present oscillatory surface waves with $n_{+}<0$ and 
$n_{-}>0$, thus implying that the `+' wave propagates in the positive 
X-direction, whereas, the `-' wave propagates in the negative X-direction. As 
the value of $u_{\rm 0o}$ is increased through
$c_{{\rm k}f}/\left(\alpha_{{\rm o}f}\right)^{1/2}$, so that
$c_{{\rm k}f}/\left(\alpha_{{\rm o}f}\right)^{1/2} < u_{\rm 0o}<c_{{\rm k}f}/
\left(\alpha_{{\rm o}f}\alpha_{{\rm e}f}\right)^{1/2}$, oscillatory surface 
modes still pertain, but now with $n_{\pm}<0$, thus implying that
both the `+' and the `-' waves now propagate in the positive X-direction. Thus,
with the increase of the slab speed  past its critical value $c_{{\rm k}f}/
\left(\alpha_{{\rm o}f}\right)^{1/2}$, the `-' surface wave reverses the 
direction of its phase propagation to be simply carried by the flow. In other 
words, the `-' surface wave changes its character from a {\em forward wave} to 
a {\em backward wave} (eg. Ostrovskii et al. 1986). To examine the energetics 
of these surface waves, we note from equation (36) that,

\begin{eqnarray}
\Delta h_{\left(\rm kink/sausage\right)}(n_{\pm}) & = & \pm n_{\pm}
\left(n_{+}-n_{-}\right)\frac{1}{k}|\tilde{\eta}^{\pm}_{
\left(\rm even/odd\right)}(k,t)|^2
\end{eqnarray}

\noindent so that,

\newcounter{lett4}
\setcounter{lett4}{38}
\setcounter{equation}{0}
\renewcommand{\theequation}{\arabic{lett4} \alph{equation}}

\begin{eqnarray}
\Delta h_{\left(\rm kink/sausage\right)}(n_{\pm})& > &0,~ \mbox {when}~ 
u_{\rm 0o} < c_{{\rm k}f}/\left(\alpha_{{\rm o}f}\right)^{1/2},
\end{eqnarray}

\noindent and

\vspace{-1.0cm}
\begin{eqnarray}
\Delta h_{\left(\rm kink/sausage\right)}(n_{+})& > &0~\mbox{and}~\Delta 
h_{\left(\rm kink/sausage\right)}(n_{-}) <0,\nonumber  \\
& &\mbox {when}~ c_{{\rm k}f}/\left(\alpha_{{\rm o}f}\alpha_{{\rm e}f}
\right)^{1/2}
> u_{\rm 0o}>c_{{\rm k}f}/\left(\alpha_{{\rm o}f}\right)^{1/2}.
\end{eqnarray}

\setcounter{equation}{38}
\renewcommand{\theequation}{\arabic{equation}}

\noindent The {\em backward} `-' surface wave is then also a {\em negative 
energy wave} in this particular reference frame - a result, that is in 
agreement with  Cairns (1979), Ostrovskii et al. (1986) and Ryutova (1988). As 
the velocity $u_{\rm 0o}$ of the slab passes through its threshold value 
$c_{{\rm k}f}/\left(\alpha_{{\rm o}f}\alpha_{{\rm e}f}\right)^{1/2}$ for 
{\em K-H instability}, an unsatble region is produced by a coalescence of the 
positive and the negative energy modes. We may note that, although the sign of 
energy of the modes depends on the choice of the co-ordinate frame, the 
existence criterion of the unstable branch (see equation (21)) is independent 
of such a choice. Also invariant is the total energy absorbed by the system 
from the external perturbation. For a $\delta$-function type perturbation, 
this energy is given by (see equation 35)

\begin{eqnarray}
\Delta h_{\left(\rm kink/sausage\right)}= \Delta h_{\left(\rm 
kink/sausage\right)}(n_{+}) + \Delta h_{\left(\rm kink/sausage\right)}(n_{-})
 =  k|\tilde{\sigma}_{\left(\rm kink/sausage\right)}(-k)|^2,
\end{eqnarray}

\noindent which is less than the excitation energy $\Delta h_{\left(\rm 
kink/sausage\right)}(n_{+})$ of the `+' mode alone in the above example. This 
extra energy is released during the process of excitaion of a negative energy 
wave, thus exciting simultaneously a positive energy wave through the mode 
interactions in the presence of the external stress $\tilde{\sigma}(-k,t)$.

\begin{center}
{\bf 4. Effects of viscosity on the wavemodes}
\end{center}

\noindent To examine the effects of {\em viscous dissipation} on the surface 
modes of the slab, we first observe that a canonical form of the stress-free 
equation of motion of the slab can be obtained by substituting 
$\partial/\partial t$ for $in$ in equation (20). This equation is

\newcounter{lett5}
\setcounter{lett5}{40}
\setcounter{equation}{0}
\renewcommand{\theequation}{\arabic{lett5} \alph{equation}}

\begin{eqnarray}
\ddot{\tilde{\eta}}_{\left(\rm kink/sausage\right)}^{\rm s}(k,t) + 
2ik\bar{U}\dot{\tilde{\eta}}_{\left(\rm kink/sausage\right)}^{\rm s}(k,t) + 
k^2\left(\delta n\right)^2\tilde{
\eta}_{\left(\rm kink/sausage\right)}^{\rm s}(k,t) = 0,
\end{eqnarray}

\noindent with

\vspace{-1.0cm}
\begin{eqnarray}
\bar{U} = \left(\alpha_{{\rm e}f}u_{\rm 0e}+\alpha_{{\rm o}f}u_{\rm 0o}\right),
\mbox{and}~\left(\delta n\right)^2 = c_{{\rm k}f}^2 - \alpha_{{\rm o}f}
\alpha_{{\rm e}f}\left(u_{\rm 0e}-u_{\rm 0o}\right)^2,
\end{eqnarray}

\setcounter{equation}{40}
\renewcommand{\theequation}{\arabic{equation}}

\noindent in a frame `s'(say), in which the two fluids `o' and `e' are seen to 
move with velocities $u_{\rm 0o}$ and $u_{\rm 0e}$, respectively. In this 
frame of reference `s', a wave profile ${\eta}^{\rm s}(x,z^{\prime},t)$ at any 
point $z=z^{\prime}$ inside the slab is seen to have a dependence given by 
${\eta}^{\rm s}(x,z^{\prime},t)={\eta}^{\rm s}(t=0)\exp{\left\{i
\left(n_{\pm}^{\rm s}t+kx\right)\right\}}$, in which the frequencies 
$n_{\pm}^{\rm s}$ have a nett drift term $k{\bar U}$, so that 
$n_{+}^{\rm s}\neq - n_{-}^{\rm s}$, signifying that the {\em reflection 
symmetry} is lost.  

Consider now a frame of refernce `r', that moves with a relative velocity 
$\bar{U}$ with respect to the `s' frame, so that the transformation 
r$\rightarrow$s is given by

\begin{eqnarray}
\tilde{\eta}^{\rm s}_{\left(\rm kink/sausage\right)}(k,t) & = & 
\tilde{\eta}^{\rm r}_{\left(\rm kink/sausage\right)}(k,t)\exp{
\left(-ik{\bar U}t\right)},
\end{eqnarray}

with $\tilde{\eta}^{\rm r}_{\left(\rm kink/sausage\right)}(k,t)$ satisfying 
the equation of a simple harmonic oscillator

\begin{eqnarray}
\ddot{\tilde{\eta}}_{\left(\rm kink/sausage\right)}^{\rm r}(k,t) + 
k^2(\delta n)^2\tilde{\eta}_{\left(\rm kink/sausage\right)}^{\rm r}(k,t)= 0.
\end{eqnarray}

\noindent A wave profile has a dependence given by 
${\eta}^{\rm r}(x,z^{\prime},t)={\eta}^{\rm r}(t=0) \exp{\left\{i
\left(n_{\pm}^{\rm r}t+kx\right)\right\}}$
in this reference frame `r', which yields $n_{+}^{\rm r} = - n_{-}^{\rm r} 
= \delta n$ (see equations (20) and (40), with $\bar U = 0$). The slab waves 
in this `r' frame thus possess a {\em reflection symmetry}, since $\delta n$ 
is an invariant that depends only on the relative velocity   
$|u_{\rm 0e}-u_{\rm 0o}|$, and not on the drift velocity $\bar U$.

While examining the effect of viscosity on the surface modes of the slab, we 
must begin our investigations by calculating the viscous dissipation as is 
seen in the reference frame `r'. This approach is in agreement with Rayleigh 
(1883), who argued that in a moving stream flowing with a velocity $\bar U$, 
the nett pressure fluctuation due to viscous drag must be 
$\delta p = \chi_{\rm d}\left(\phi-\bar{U}x\right)$, as measured in the frame 
of reflection symmetry of the perturbations, with $\chi_{\rm d}$ being a 
viscous drag coeffecient and $\phi (x,z,t)$ being the velocity potential. In 
the present case of the magnetic modes of the slab, this requires that the 
rate of mechanical energy dissipation of the slab due to viscous damping (eg. 
Landau and Lifshitz 1959a) is given by

\newcounter{lett6}
\setcounter{lett6}{43}
\setcounter{equation}{0}
\renewcommand{\theequation}{\arabic{lett6} \alph{equation}}

\begin{eqnarray}
{\dot{\cal E}_{\nu}}^{\rm r} & = & -\int\!\left[\int_{-\infty}^{-d}\Psi \,dz
+ \int_{d}^{\infty}\Psi \,dz\right]dx,
\end{eqnarray}

\noindent where,

\vspace{-1.0cm}
\begin{eqnarray}
\Psi = 2\rho_{\rm 0e}\nu_{\rm e}\left[\left(\frac{\partial}{\partial x}
\left(\delta u_{x{\rm e}}\right)^{\rm r}\right)^2 + \left(\frac{\partial}
{\partial z}\left(\delta u_{x{\rm e}}\right)^{\rm r}\right)^2
+ \left(\frac{\partial}{\partial z}\left(\delta u_{z{\rm e}}\right)^{\rm r}
\right)^2\right],
\end{eqnarray}

\setcounter{equation}{43}
\renewcommand{\theequation}{\arabic{equation}}

\noindent with $\nu_{\rm e}$ being the kinematic viscosity in medium `e', 
whereas, medium `o' is taken to be inviscid. It is possible to use the 
classical gas dynamical formula for viscous dissipation (as given in equation 
(43)), while retaining the velocity discontinuities at the slab interfaces, 
only in such a situation, where either the internal or the external medium 
alone has viscous dissipation and the other medium is inviscid, ie., the 
details arising due to boundary layer may be ignored and the tangential 
discontinuity of the velocities at the interfaces still remains a valid 
condition. We however note that, ideally one should consider an anisotropic 
viscous stress tensor in the presence of a magnetic field (cf. Braginskii 
1965), as in the situation considered here. Whatever the case may be, the 
specific choice of the viscous stress tensor is not expected to change the 
overall stability properties of the modes about which we are mainly concerned 
in this paper.

We now substitute the expressions

\newcounter{lett7}
\setcounter{lett7}{44}
\setcounter{equation}{0}
\renewcommand{\theequation}{\arabic{lett7} \alph{equation}}

\begin{eqnarray}
\delta u_{x{\rm e}}^{\rm r}(x,z,t) & = & \mp k \dot{\tilde{\eta}}^{\rm r}
(k,\mp d,t)\exp{\left\{\pm k\left(z\pm d\right)\right\}}\exp{\left(ikx\right)} 
                             \nonumber \\
& = & \mp \left(\delta n\right)\tilde{\eta}^{\rm r}(k,\mp d,t)\exp{
\left\{\pm k\left(z\pm d\right)\right\}}\exp{\left(ikx\right)},
\end{eqnarray}

\noindent and

\vspace{-1.0cm}
\begin{eqnarray}
\delta u_{z{\rm e}}^{\rm r}(x,z,t) & = & \dot{\tilde{\eta}}^{\rm r}
(k,\mp d,t)\exp{\left\{\pm\left(z\pm d\right)\right\}}\exp{\left(ikx\right)} 
                                                                \nonumber \\
& = & \left(\delta n\right)\tilde{\eta}^{\rm r}(k,\mp d,t)\exp
{\left\{\pm k\left(z\pm d\right)\right\}}\exp{\left(ikx\right)},
\end{eqnarray}

\setcounter{equation}{44}
\renewcommand{\theequation}{\arabic{equation}}

\noindent for the various perturbations in equation (43). Using the 
definitions given in equation (3, see Section 2.1), we thus obtain

\begin{eqnarray}
{\dot{\cal E}_{\nu}}^{\rm r}\left(\tilde{\eta}_{\left(\rm kink/sausage
\right)}^{\rm r}(k,t),\dot{\tilde{\eta}}_{\left(\rm kink/sausage
\right)}^{\rm r}(k,t)\right) & = & -2i\rho_{\rm 0e}\nu_{\rm e}k(\delta n)
\left[\tilde{\eta}_{\left(\rm kink\right)}^{\rm r}(k,t)\dot{
\tilde{\eta}}_{\left(\rm kink\right)}^{\rm r}(k,t)\right. \nonumber \\
& & \left. + \tilde{\eta}_{\left(\rm sausage\right)}^{\rm r}(k,t)\dot
{\tilde{\eta}}_{\left(\rm sausage\right)}^{\rm r}(k,t)\right],
\end{eqnarray}

\noindent for the rate of viscous dissipation in frame `r'. 

The energy thus dissipated in frame `r' gives rise to an increament $\delta 
S_{\rm 0e}$ in the entropy of the system, that must be invariant in all 
frames. Noting that, $\delta n$ in equation (45) is an invariant (see, 
equation (40b)), we obtain the rate of increase of entropy (or, the heating 
rate) in terms of the quantities defined in the frame `s'. Thus, substituting 
$\delta n = n+k{\bar U}$, we have

\begin{eqnarray}
T_{\rm 0e}\frac{\delta S_{\rm 0e}}{\delta t}\left(\tilde{\eta}_{\left(\rm 
kink/sausage\right)}^{\rm s}(k,t),\dot{\tilde{\eta}}_{\left(\rm 
kink/sausage\right)}^{\rm s}(k,t)\right) & = & 2i\rho_{\rm 0e}\nu_{\rm e}
k(n+k\bar U) \left[\tilde{\eta}_{\left(\rm kink\right)}^{\rm s}(k,t)\dot{
\tilde{\eta}}_{\left(\rm kink\right)}^{\rm s}(k,t)\right. \nonumber \\
& & \left. + \tilde{\eta}_{\left(\rm sausage\right)}^{\rm s}(k,t)\dot
{\tilde{\eta}}_{\left(\rm sausage\right)}^{\rm s}(k,t)\right],
\end{eqnarray}

\noindent where, $T_{0e}$ is the equlibrium temperature of the medium `e'.

The heating rate being thus known, we demand that, the {\em thermodynamic 
potential} $\Phi_{\rm 0}$ ($= H$$-T_{\rm 0e}S_{\rm 0e})$  must be minimum at 
all instants for the wave propagation to be a manifestation of the system's 
response to its departure from equilibrium (cf. Landau and Lifshitz 1959b; 
Glansdorff and Prigogine 1971). This shows,

\begin{eqnarray}
& & \delta \Phi_{\rm 0}(t) \equiv  \Phi_{\rm 0}(t+\delta t)-\Phi_{\rm 0}(t) 
                                                            \nonumber \\
& & ~~~~~~~= ~~~ \left\{\frac{\partial H}{\partial \dot{\tilde{\eta}}_{
\left(\rm kink/sausage\right)}}\ddot{\tilde{\eta}}_{\left(\rm kink/sausage
\right)} + \frac{\partial H}{\partial \tilde{\eta}_{\left(\rm 
kink/sausage\right)}}\dot{\tilde{\eta}}_{\left(\rm kink/sausage\right)}-
T_{\rm 0e}\frac{\delta S_{\rm 0e}}{\delta t}\right\}\delta t = 0, \nonumber \\
& & 
\end{eqnarray}

\noindent for any infinitesimal $\delta t$. In equation (47), we have dropped 
the superscript `s', still indicating the observer's frame.  With the help of 
equation (18), we then obtain (after substituting  $\tilde{\sigma}_{\left(\rm 
kink/sausage\right)}(-k,t) = 0$),

\begin{eqnarray}
& &\left\{\frac{k^2B_{0e}^2}{16\pi n^2} + \frac{k^2B_{0o}^2}{8\pi n^2}
\left[\frac{2\sinh{(2kd)}\pm 4(kd)}{8\cosh^2{(kd)}}\right]\right\}
\ddot{\tilde{\eta}}_{\left(\rm kink/sausage\right)}(k,t)\dot{
\tilde{\eta}}_{\left(\rm kink/sausage\right)}(k,t) \nonumber \\
& &+\left\{\left[\rho_{\rm 0e}\left(n+ku_{\rm 0e}\right)^2 - 
\frac{3k^2B_{\rm 0e}^2}{16 \pi}\right] 
+ \left[\rho_{\rm 0o}\left(n+ku_{\rm 0o}\right)^2 - \frac{k^2B_{\rm 0o}^2}
{4\pi}\right]f(k)\right. \nonumber \\
& & \left. + \frac{k^2B_{\rm 0o}^2}{8\pi}\left[
\frac{2\sinh{(2kd)}\pm 4(kd)}{8\cosh^2{(kd)}}\right]\right\}
\tilde{\eta}_{\left(\rm kink/sausage\right)}(k,t)\dot{\tilde{\eta}}_{\left(\rm 
kink/sausage\right)}(k,t)\nonumber \\
& &-\rho_{\rm 0e}\nu_{e}k^2\left(\frac{n+k\bar{U}}{n}\right)\dot
{\tilde{\eta}}_{\left(\rm kink/sausage\right)}^2(k,t) = 0,
\end{eqnarray}

\noindent which is true for all $\dot{\tilde{\eta}}_{\left(\rm 
kink/sausage\right)}(k,t)$. Substituting ${\partial}/{\partial t} = in$, and 
also requiring a non-trivial solution, we then obtain the dispersion relation

\begin{eqnarray}
\rho_{\rm 0e}\left[\left(n+ku_{\rm 0e}\right)^2-k^2c_{\rm Ae}\right] + 
\rho_{\rm 0o}\left[\left(n+ku_{\rm 0o}\right)^2-k^2c_{\rm Ao}\right]f(k) = 
i\rho_{\rm 0e}\nu_{\rm e}k^2(n+k\bar U),
\end{eqnarray}

\noindent for the viscous surface modes of the slab. Note the factor 
$(n+k{\bar U})$ in the damping term of equation (49). This factor 
differentiates equation (49) from the earlier results (cf. Kikina 1967; 
Weissman 1970; Cairns 1979; Ezerskii et al. 1981; Ostrovskii and Stepanyants 
1982; Ostrovskii et al. 1986; Ruderman and Goossens 1995), in which the 
viscous drag force was  proportional to the frequency $n$ of the waves in the 
observer's frame `s', thus depending on the velocity $\bar U$ of the material.
In view of this important difference, it is here pertinent, that we discuss
the significance of equation (49) in some detail.

The flow of the two fluids creates a momentum flux $\rho_{\rm 0o}u_{\rm 0o}+
\rho_{\rm 0e}u_{\rm 0e}$ per unit volume in the observer's (`s') frame, that 
is equivalent to imposing a velocity ${\bar U}=\left(\alpha_{{\rm o}f}u_{
\rm 0o}+\alpha_{{\rm e}f}u_{\rm 0e}\right)$ on all matter in the wave profile. 
Moving to  any other frame, where velocities of the fluids are $u_{\rm 0o}^{
\prime}= u_{\rm 0o}+u$ and $u_{\rm 0e}^{\prime}=u_{\rm 0e}+u$, we have 
${\bar U}^{\prime}= \alpha_{{\rm o}f}u_{\rm 0o}^{\prime}+\alpha_{{\rm e}f}u_{
\rm 0e}^{\prime}={\bar U}+u$. This nett velocity of the wave profile appears 
purely due to Galilean transformation, and should not contribute to any 
process of exchange of energy or momentum within the system, and thus cannot 
contribute to dissipation. Contrary to the earlier results, the expression for 
the viscous drag must, therefore, have no explicit dependence on the drift 
velocity $\bar U$, as is evident by the appearance of the invariant factor 
$\delta n = n+k{\bar U}$ in the damping force in equation (49).

Returning to the modes of oscillations of the magnetic slab, equation (49) 
yields solutions

\begin{eqnarray}
n_{\pm} = \frac{i}{2}\alpha_{{\rm e}f}\nu_{\rm e}k^2 - k\left[{\bar U}\pm 
\left\{\left[c_{{\rm k}f}^2 - \alpha_{{\rm e}f}\alpha_{{\rm o}f}
\left(u_{\rm 0e}-u_{\rm 0o}\right)^2\right]
-\frac{1}{4}\alpha_{{\rm e}f}^2\nu_{\rm e}^2k^2\right\}^{1/2}\right],
\end{eqnarray}

\noindent thus showing that, for flow velocities below the threshold for 
the K-H instability, the principal effect of viscous dissipation is to 
introduce a damping for both the positive and the negative energy modes of 
the magnetic slab - a result, that is in contradiction to the earlier results 
(see the references above), which  predict  a {\em dissipative instability} 
for the negative energy waves. The main consequence of our considering the 
correct Galilean transformation, while examining the viscous effects on the 
slab waves, is then the establishment of the fact that, the stability property 
of the modes remains {\em non-singular} in the presence of a small dissipation,
that changes only slightly the threshold for the K-H instability of the slab, 
with the modified instability criterion given by

\begin{eqnarray}
|u_{\rm 0e}-u_{\rm 0o}| \geq \left[\frac{c_{{\rm k}f}^2}{\alpha_{{\rm e}f}
\alpha_{{\rm o}f}}- \frac{1}{4}\left(\frac{\alpha_{{\rm e}f}}{
\alpha_{{\rm o}f}}\right)\nu_{\rm e}^2k^2\right]^{1/2},
\end{eqnarray}

\noindent that smoothly approaches the adiabatic criterion in equation (21) for 
a vanishingly small kinematic viscosity $\nu_{\rm e}\rightarrow 0$.

\newpage
\begin{center}
{\bf 5. Concluding remarks}
\end{center}

\noindent Occurrence of magnetic structures is abundant in various 
astrophysical situations, such as in the solar photospheric flux tubes, or in 
the solar coronal plasma loops.  Such magnetic structures are often associated 
with field-aligned plasma flows, with the velocities of these flows being 
different inside the structures than those outside, thus producing 
{\em shearing motions} in the plasma medium. Detailed understandings of the 
complex interactions of such shearing flows with the oscillatory motions of 
the structures are necessary to study accurately the energy transport 
processes in astrophysics, such as the mechanisms of non-thermal energy 
transport from the solar sub-surface layers to the upper solar atmosphere. 
Certain investigations have been carried out (eg. Ryutova 1988; Nakariakov and 
Roberts 1995; Nakariakov et al. 1996; Ruderman and Goossens 1995; Ruderman et 
al. 1996; Joarder et al. 1997) in this direction, that highlited the role of 
negative energy waves in such processes. As a further contribution to such 
investigations, we here examined in detail certain specific aspects of the 
interactions of magnetohydrodynamic waves with shearing flows, and particularly 
of the negative energy waves, by using a self-consistent thermodynamic 
approach. This approach helped us to generalize the expression for the 
hydrodynamic wave energy given in Cairns (1979) to magnetohydrodynamics (see, 
equation (36) in Section 3.3), thus enabling us to  calculate the energy
of the hydromagnetic  waves (of course in the harmonic approximation), when 
the linear dispersion relations of such waves are known along with the 
observationally obtained informations regarding the wave amplitudes. Once the 
wave energy is thus calculated, equation (35) then guides us to obtain a 
rough estimate of the generating stresses $\tilde{\sigma}(-k,t)$ of the waves.
Such estimates of the stresses may be of great importance in several 
astrophysical situations, particularly in solar MHD cases, where such 
estimates may provide us with some clues regrading the physical processes 
that may be taking place in the sub-surface layers of the Sun, or in the 
regions of complex magnetic topology in the solar atmosphere, about which we 
have very little direct observational evidence. Finally, by incorporating 
viscosity, we obtain the dispersion relations (equation 49 in Section 4) 
which, while precluding the possibility of dissipative instability,- sets the 
correct conditions for the stability of the system (equation (51)) and also 
yields the time scales for the decay of the disturbances in the surface modes 
of MHD systems. It is to be hoped, that the present study would provide us 
with some guidance in gaining further physical insights into the complex 
nature of the interactions between the magnetic field and the fluid flows in 
various astrophysical systems,- both for estimates in terms of energetics as 
also in the study of evolutions of MHD eigenmodes. 

\begin{center}
{\bf 6. Acknowledgements.}
\end{center}

\noindent One of us (PSJ) is indebted to the members of the solar theory 
group of the University of St. Andrews, and particularly to Professor 
B. Roberts and Dr. V. M. Nakariakov for the inspirations, constant 
encouragements, expertise and the warm hospitality
that he received during his stay as a PPARC visiting fellow in that 
University. Dr. Nakariakov initiated him to the topic of negative energy 
waves. Discussions with Professors R. A. Cairns and A.D.D Craik on this topic 
are most gratefully acknowledged.

\newpage
\begin{center}
{\bf References.}
\end{center}

\noindent Acheson, D. J. 1976 {\em J. Fluid Mech.} {\bf 77}, 433.

\noindent Alfven, H. 1950 {\em Cosmical Electrodynamics.} Clarendon Press, 
Oxford.

\noindent Benjamin, T. B. 1963 {\em J. Fluid Mech.} {\bf 16}, 436.

\noindent Braginskii, S. I. 1965 in {\em Rev. Plasma Phys.}(ed. M. A. 
Leontovich), {\bf I}, p. 205.

\noindent Cairns, R. A. 1979 {\em J. Fluid Mech.} {\bf 92}, 1.

\noindent Chandrasekhar, S. 1961 {\em Hydrodynamic and Hydromagnetic 
 Stability.} Clarendon Press, Oxford.

\noindent Cowling, T. G. 1957 {\em Magnetohydrodynamics.} Wiley-Interscience, 
New York.

\noindent Craik, A. D. D. 1985 {\em Wave Interaction and Fluid Flows.} 
Cambridge Univ. Press, Cambridge.

\noindent Craik, A. D. D. and Adam, J. A. 1979 {\em J. Fluid Mech.} {\bf 92}, 
15.

\noindent Edwin, P. M. and Roberts, B. 1982 {\em Solar Phys.} {\bf 76}, 239.

\noindent Edwin, P. M. and Roberts, B. 1983 {\em Solar Phys.} {\bf 88}, 179.

\noindent Ezerskii, A. B., Ostrovskii, L. A. and Stepanyants, Yu. A. 1981 
{\em Izv. Atmos. Ocean. Phys.} {\bf 17}, 890.

\noindent Glansdorff, P. and Prigogine, I. 1971 {\em Thermodynamic Theory of 
Structure, Stability and Fluctuations.} Wiely-Interscience, New York.

\noindent Joarder, P. S., Nakariakov, V. M. and Roberts, B. 1997 {\em Solar 
Phys.} (in Press).

\noindent Kikina, N. G. 1967 {\em Sov. Phys. Akust.} {\bf 13}, 184.

\noindent Landau, L. D. and Lifshitz, E. M. 1959a {\em Fluid Mechanics.} 
Pergamon Press, Oxford.

\noindent Landau, L. D. and Lifshitz, E. M. 1959b {\em Statistical Physics.}, 
Part I. Pergamon Press, Oxford.

\noindent Nakariakov, V. M. and Roberts, B. 1995 {\em Solar Phys.} {\bf 159}, 
213.

\noindent Nakariakov, V. M., Roberts, B. and Mann, G. 1996 {\em Astron. 
Astrophys.}, {\bf 311}, 311.

\noindent Ostrovskii, L. A. and Stepanyants, Yu. A. 1982 {\em Izv. Akad. 
Nauk SSSR.} Ser Mekh. Zheidk. Gaza No. 4, 63.

\noindent Ostrovskii, L. A., Rybak, S. A. and Tsimring, L. Sh. 1986 
{\em Sov. Phys. Usp.} {\bf 29}, 1040.

\noindent Rayleigh, Lord. 1883 {\em Proc. Lon. Math. Soc.} {\bf XV}, 69.

\noindent Roberts, B. 1981a {\em Solar. Phys.} {\bf 69}, 27.

\noindent Roberts, B. 1981b {\em Solar. Phys.} {\bf 69}, 39.
 
\noindent Ruderman, M. S. and Goossens, M. 1995 {\em J. Plasma Phys.} 
{\bf 54}, 149.

\noindent Ruderman, M. S., Verwichte, E., Erdelyi, R. and Goossens, M. 1996 
{\em J. Plasma Phys.} {\bf 56}, 285.

\noindent Ryutova, M. P. 1988 {\em Sov. Phys. JETP.} {\bf 67}, 1594.

\noindent Satya Narayanan, A. 1991 {\em Plasma Phys. Control. Fusion} 
 {\bf 33}, 333.

\noindent Singh, A. P. and Talwar, S. P. 1994 {\em Solar Phys.} 
{\bf 149}, 331.

\noindent Sommerfeld, A. 1950 {\em Mechanics of Deformable bodies.} Academic 
Press, New York.

\noindent Stix, T. H. 1962 {\em The Physics of Plasma Waves.} Mc.Graw-Hill, 
New York.

\noindent Weissman, M. A. 1970 {\em Notes on Summer Study Prog. Geophys. 
Fluid Dyn.} Woods Hole Oceanog. Inst. no. 70-50.

\noindent Witham, G. B. 1974 {\em Linear and Non-Linear Waves.} 
Wiley-Interscience, New-York.

\newcounter{alett}
\setcounter{alett}{1}
\setcounter{equation}{0}
\renewcommand{\theequation}{\Alph{alett}.\arabic{equation}}

\newpage
\begin{center}
{\bf Appendix. Energy density of an incompressible MHD plasma.}
\end{center}

\noindent The {\em energy density} of an incompressible magnetohydrodynamic 
system is defined as

\begin{eqnarray}
\varepsilon \left({\boldmath r},t\right) & = & \varepsilon_{\rm i}
\left({\boldmath r},t\right) + \frac{B^2\left({\boldmath r},t\right)}{8\pi} + 
\frac{\rho_{\rm 0}}{2}u^2\left({\boldmath r},t\right),
\end{eqnarray}

\noindent where, $\varepsilon_{\rm i}$ is the {\em thermodynamic internal 
energy} of the plasma, and the quantities $\rho_{\rm 0}$, $u$ and $B$ are as 
defined in equation (2) of the main text.

In the present case, we consider the fluid to be {\em incompressible}, and the 
hydrodynamic processes to be {\em adiabatic}. In that case, we write (cf. 
Landau and Lifshitz, 1959b)

\begin{eqnarray}
\varepsilon_{\rm i}\left({\boldmath r},t\right) & = & \mu \left({\boldmath r},
t\right)/\upsilon,
\end{eqnarray}

\noindent where, $\mu \left({\boldmath r},t\right)$ is the local {\em chemical 
potential} of the system, and $\upsilon$ is its {\em specific volume}. Thus, 
for any fluctuation
in the thermodynamic state of the system, we have

\begin{eqnarray}
\delta \varepsilon_{\rm i}\left({\boldmath r},t\right) & = & \delta \mu 
\left({\boldmath r},t\right)/\upsilon,
\end{eqnarray}

\noindent where, $\upsilon$ is a constant in an incompressible fluid. Further,

\begin{eqnarray}
\delta \mu & = & - s\,dT + \upsilon\,dp,
\end{eqnarray}

\noindent where, $s$ is the {\em specific entropy} of the system.

If we now assume the {\em electrical} and the {\em thermal conductivities} of 
the fluid to be infinite, and its {\em viscosity coefficient} to be zero, then 
the system cannot support any thermal gradients, and, therefore, $\delta T$ is 
zero at all points. Combining equations (A.2-A.4), we then find that the 
fluctuations in the thermodynamic energy density to be (see, Sommerfeld 1950 
for an alternative interpretation)

\begin{eqnarray}
\delta \varepsilon_{\rm i} & = & \delta p,
\end{eqnarray}

\noindent so that, neglecting the integration constant, we ultimately obtain 
equation (6, Section 2.2) of the main text, ie.,

\begin{eqnarray}
\varepsilon \left({\boldmath r},t\right) = p\left({\boldmath r},t\right) + \frac{B^2\left(\boldmath r,t\right)}{8\pi} + \frac{\rho_{\rm 0}}{2}u^2\left(\boldmath r,t\right)
\end{eqnarray}

for an {\em incompressible} fluid.

\end{document}